\documentclass{aa}  

\usepackage{graphicx}
\usepackage{txfonts}
\usepackage{breqn}
\usepackage{longtable}
\usepackage{natbib}
\usepackage{ulem} 
\usepackage{float}
\newcommand{\mygi}{MyGIsFOS}

\newcommand{\logg}{\ensuremath{\log\,g}}

\def\teff{$T\rm_{eff}$}
\newcommand{\kms}{$\rm km s ^{-1}$}

\begin{document} 

\title{The Pristine survey\\ XXVI. Chemical abundances of subgiant stars of the extremely
metal-poor stream C-19\thanks{Based on observations made at ESO with X-shooter, programme 
109.231F.001} 
}
\titlerunning{Pristine XXIV - C19 SG}

\author{
P.~Bonifacio \inst{1} \and
E.~Caffau    \inst{1} \and
P.~Fran\c{c}ois \inst{2,3} \and
N.~Martin \inst{4} \and
R.~Ibata \inst{4} \and
Z.~Yuan \inst{4} \and
G.~Kordopatis \inst{5} \and
J.~I.~Gonz\'alez~Hern\'andez \inst{6,7} \and
D.~S.~Aguado \inst{6,7} \and 
G.~F.~Thomas \inst{6,7} \and A. Viswanathan\inst{8}\and 
E.~Dodd \inst{8} \and 
F.~Gran \inst{5} \and
E. Starkenburg \inst{8} \and
C.~Lardo \inst{9,10}\and 
R.~Errani \inst{11}\and
M.~Fouesneau\inst{12}\and
J. F. Navarro\inst{13}\and
K. A. Venn\inst{13}\and
K.~Malhan\inst{14}
} 

\institute{GEPI, Observatoire de Paris, Universit\'{e} PSL, CNRS,  5 Place Jules Janssen, 92190 Meudon, France
\and
 GEPI, Observatoire de Paris, Universit\'{e} PSL, CNRS,  77 Av. Denfert-Rochereau, 75014 Paris, France
\and
UPJV, Universit\'e de Picardie Jules Verne, 33 rue St Leu, 80080 Amiens, France
\and
Universit{\'e} de Strasbourg, CNRS, Observatoire astronomique de Strasbourg, UMR 7550, F-67000 Strasbourg, France
\and
Universit{\'w} C{\^o}te d'Azur, Observatoire de la C{\^o}te d'Azur, CNRS, Laboratoire Lagrange, 06304, Nice, France
\and
 Instituto de Astrof\'{i}sica de Canarias, E-38205 La Laguna, Tenerife, Spain
\and
Departamento de Astrof\'{i}sica, Universidad de La Laguna, E-38206 La Laguna, Tenerife, Spain
\and 
Kapteyn Astronomical Institute, University of Groningen, Landleven 12, 9747 AD Groningen, The Netherlands
\and
Dipartimento di Fisica e Astronomia, Universita degli Studi di Bologna, Via Gobetti 93/2, I-40129 Bologna, Italy
\and
INAF - Astrophysics and Space Science Observatory of Bologna, Via Gobetti 93/3, 40129, Bologna, Italy
\and
McWilliams Center for Cosmology, Department of Physics, Carnegie Mellon University, Pittsburgh, PA 15213 USA;
\and
Max Plank Institute for Astronomy (MPIA), K{\"o}nigstuhl 17, 69117 Heidelberg, Germany.
\and
Department of Physics and Astronomy, University of Victoria, Victoria, BC, Canada V8P 5C2.
\and
DARK, Niels Bohr Institute, University of Copenhagen, Jagtvej 128, 2200 Copenhagen, Denmark
}

   \date{Received July 15, 20204; accepted December 14, 2024}

  \abstract
{ The C-19 stellar stream is the most metal-poor stream known to date. While its width and velocity dispersion indicate a dwarf galaxy origin, its metallicity spread and abundance patterns are more similar to those of globular clusters (GCs). If it is indeed of GC origin,  its extremely low metallicity ([Fe/H]=--3.4, estimated from giant stars)
implies that these stellar systems can form out of gas that is as extremely poor in metals as this. Previously,
only giant stream stars were observed spectroscopically, although the majority
of stream stars are unevolved stars.}
{We pushed the spectroscopic observations to the subgiant branch stars
($G\approx 20$) in order to consolidate the chemical and dynamical properties 
of C-19.} 
{We used the high-efficiency spectrograph X-shooter fed by
the ESO 8.2\,m VLT telescope to observe 15 candidate subgiant C-19 members.
The spectra were used to measure radial velocities and to determine
chemical abundances using the \mygi\ code. }
{We developed a likelihood model that takes metallicity and radial velocities into account. We conclude that 
12 stars are likely members of C-19, while 3 stars
(S05, S12, and S13) are likely contaminants.
When these 3 stars are excluded, our model 
implies a mean metallicity
$\rm \langle [Fe/H]\rangle = -3.1\pm 0.1$, the  mean radial velocity is
$\langle v_r\rangle=-192\pm3$\,\kms,  and
the velocity dispersion is $\sigma_{vr}=5.9^{+3.6}_{-5.9}$\,\kms . This all agrees within errors with previous studies. 
 The A(Mg) of a sample of 15 C-19 members, including 6  giant stars, shows a  standard deviation of 0.44\,dex, and the mean uncertainty on Mg is 0.25\,dex.}
{Our preferred interpretation of the current data is that C-19 is a disrupted GC.
We cannot completely rule out  the possibility that the GC could have belonged to a dwarf galaxy that contained more metal-rich  stars, however. This  scenario would explain
 the radial velocity members at higher metallicity, as  well as the width and velocity dispersion of the stream. In either case, a GC formed out of gas as poor in metals as these stars seems necessary to explain the existence of C-19.
  The possibility that no GC was associated with C-19 cannot be ruled out either.}

   \maketitle
%
\section{Introduction\label{intro}}
  
Stream C-19 was discovered by \citet{ibata21}. Based on more spectra, some
of which have a high resolution, \citet{martin22}  interpreted it as a 
disrupted GC because of its very low metallicity dispersion and for a spread 
in Na abundances that is consistent with what is observed in bound GCs \citep{gratton01}, but not in dwarf galaxies 
\citep[see e.g.][and references therein]{simon,mcconnachie} or 
among Galactic halo field stars \citep[see e.g.][and references therein]{matteucci}. C-19 is exceptional through its mean metallicity ([Fe/H]=--3.4), which is more metal poor by an order of magnitude than the most metal-poor GC  known \citep{harris}.
The formation of GCs is only poorly understood. Although many theories
have been put forward \citep[see e.g.][and references therein]{madau20}, none has gained widespread consensus.
The fact that no Galactic GC is observed below $\rm[Fe/H]=-2.5$ was even
interpreted as a floor below which GCs cannot form \citep[see e.g.][]{kruij19, beasley2019}.
If C-19 is indeed a disrupted GC, 
this notion has to be abandoned or the level of the floor has to be pushed back
by an order of magnitude. 

While the abundance variations and metallicity spread of C-19 are consistent with a 
GC origin, its width (${\sim}200\,\mathrm{pc}$) and velocity dispersion (${\sim}6\,\mathrm{km\,s}^{-1}$) are more naturally reproduced if the progenitor contained substantial amounts of dark matter, similar to a dwarf galaxy \citep{pristineXVIII}.

All the C-19 stars that were studied spectroscopically so far (\citealt{martin22,pristineXVII,viswanathan2024}, Venn et al. 2025 in prep.)
have been giants. In order to gain further insight into the nature of C-19 and to constrain
its age from the study of the colour-magnitude diagram of its confirmed members,
we decided to push the observations to the subgiant branch, which is at
magnitude $G\approx 20$. To do this, we used the high-efficiency spectrograph X-shooter
on the ESO 8.2\,m 
In this paper, we describe the analysis and the results of these observations.

\begin{figure*}
\centering
\resizebox{!}{24truecm}{\includegraphics[clip=true]{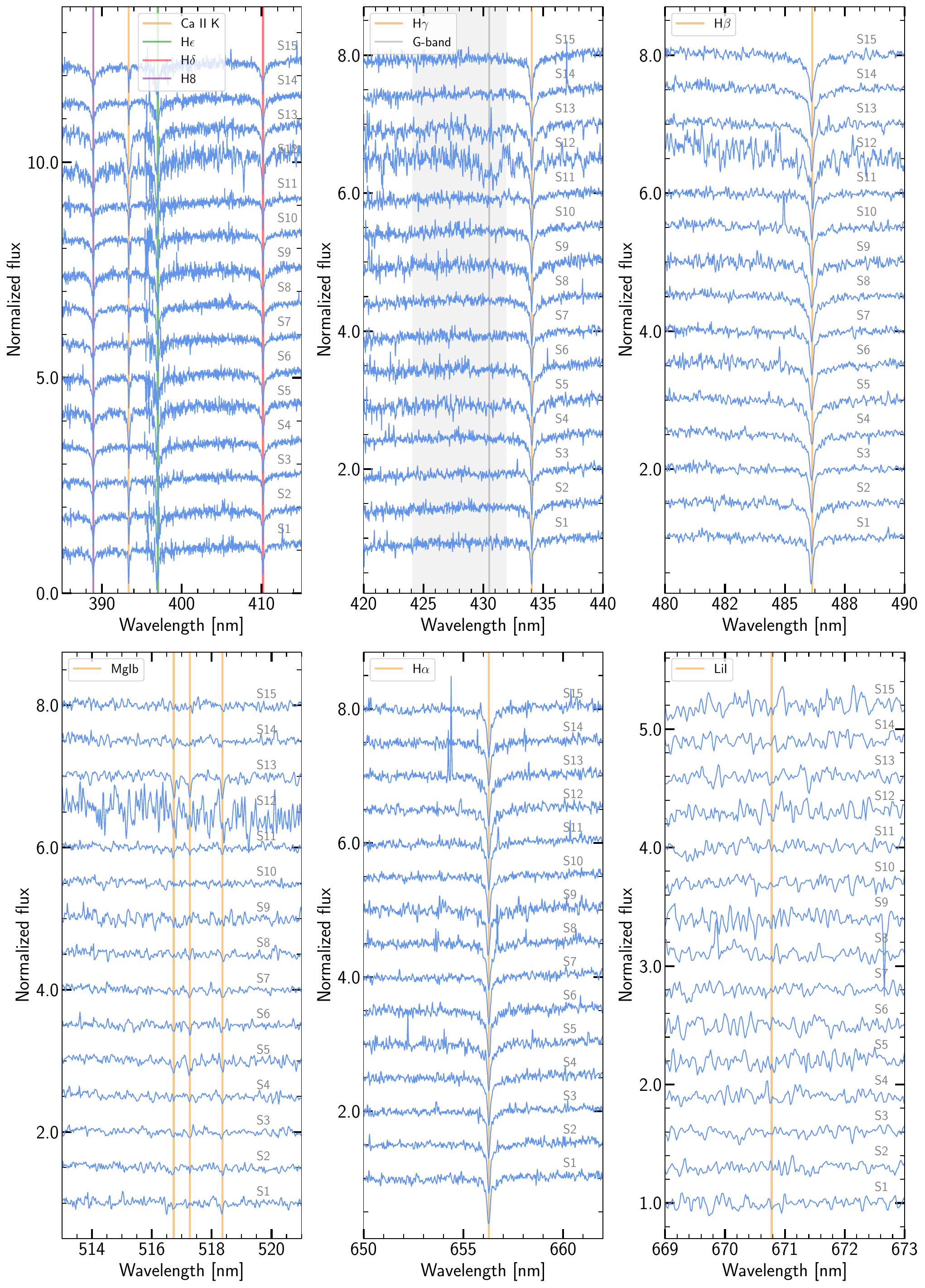}}
\caption{X-shooter spectra of the targets in several spectral regions of the UBV and VIS arms.}
\label{spec}
\end{figure*}

\section{Target selection, observations, and data reduction}

\citet{martin22} selected candidate members in the core of the C-19 stellar stream
using the algorithm {\tt STREAMFINDER} \citep{ibata21}.
The brightest candidates on the red giant branch 
were observed, and the observations were discussed
in \citet{martin22}. In Fig. 1 of the same paper, a set of candidates 
among subgiant and turn-off was presented as well.
Our targets
span the range $19.5 \le G \le 19.7$. We estimated that we could obtain a useful spectrum in this magnitude range in a one-hour
observation with X-shooter \citep{xshooter} at the VLT.
The 15 selected targets were observed in service mode between
June 8, 2022, and October 17, 2022. Most observations
had A quality, and only three had B quality.
Each star was observed for a one-hour observing block in stare mode, which implies
an integration time of about 3100\,s.
We used a 1\farcs{3} slit in the UVB arm, 1\farcs{5} slit in the VIS arm,
and 5\farcs{0} slit in the NIR arm.  The signal-to-noise ratios (S/N) of all the NIR spectra were of too
to be scientifically exploited, and the spectra are not discussed further.
The UVB and VIS CCDs were binned $2\times 2$. Because the slits we used were wide, this  did not degrade the resolving power. 
The resolving power in the two arms is 4100\ in the UVB and 5000\ in the VIS. 
The signal-to-noise ratio of our spectra ranged from 15 to 50 at 487\,nm, with a mean of 38. 
The spectra were reduced by us using the X-shooter pipeline \citep{xsh_pipeline}. In Fig.~\ref{spec} we show several spectral regions of the UBV and VIS arms. 
The \ion{Ca}{ii}~K 393.3~nm line and the \ion{Mg}{i}b 
triplet lines are clearly visible, in particular, the \ion{Mg}{i}b 518.3~nm line. The S12 and 
S13 targets are apparently significantly richer in metals, from the \ion{Ca}{ii}~K feature, and S05 also 
show a stronger \ion{Ca}{ii}~feature than the other targets. The carbon band (G-band) at 
430~nm appears to be difficult to detect 
because most of the targets are not enhanced in C. However, S12 seems to have a stronger G band. Finally, the S/N together with the resolving power make it difficult 
to detect the \ion{Li}{i} doublet feature.

 \section{Data analysis}

\begin{figure}
\resizebox{\hsize}{!}{\includegraphics[clip=true]{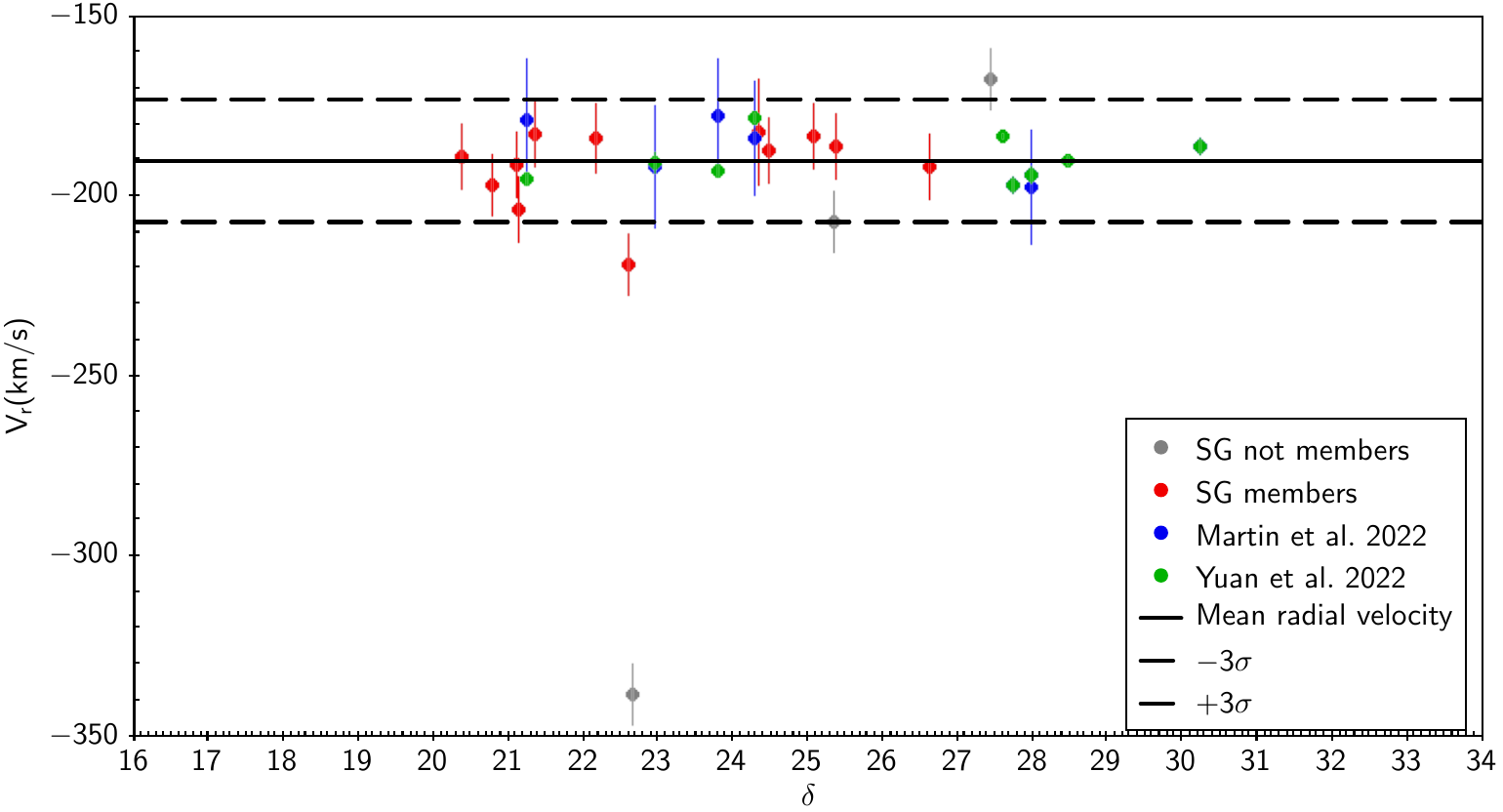}}
\caption{Radial velocities of our program stars and the stars from \citet{martin22} and 
\citet{pristineXVII} plotted as a function of declination. 
The dashed lines indicate three 
$\sigma$ shifts from the mean value (represented by the thick line).}
\label{vrad}
\end{figure}

For each spectrum, we determined the radial velocity from the VIS spectrum. 
We used our own code for the template matching, using a stellar synthetic
spectrum with parameters close to those of each star as a template. This was extracted from the synthetic
grid that we used for the abundance analysis. In addition, we used a template for the 
telluric absorption lines computed by the TAPAS service
\citep{tapas}\footnote{\url{https://tapas.aeris-data.fr/en/home/}}.
The observed spectrum was normalised using a spline through interactively defined
continuum points. The normalised stellar spectrum and telluric absorption were then allowed
to shift, each independent of the other, in order to minimise the $\chi^2$ function.
The observed radial velocity was then obtained by subtracting the velocity of the telluric
lines from the velocity of the stellar lines. The barycentric radial velocity of the star was
finally obtained by adding the barycentric correction to the observed radial velocity.
The radial velocities are provided in Table\,\ref{table:idphotvr}.
The formal error on the fit is smaller than 1.5\,\kms for all stars
except for S09, for which it is 7.3\,\kms. The pipeline
corrects the wavelength scale for  the instrument flexures, and our use of telluric lines should 
correct for the centring of the star on the slit.  According the X-shooter 
manual\footnote{\url{https://www.eso.org/sci/facilities/paranal/instruments/xshooter/doc/VLT-MAN-ESO-14650-4942_P104v1.pdf}},
the systematic wavelength accuracy is 
 7.5\,\kms\ in the VIS arm. We  estimated 
the total error on our measurements by adding this
value linearly to the statistical error.

We note that when it is cross-correlated with a synthetic template, star S09 shows an asymmetry in its correlation function for the UVB and VIS spectra. This is reflected in its large velocity uncertainty. We suspect that this star is an SB2 binary, and our derived abundances should therefore also be considered with caution for this star.

We determined atmospheric parameters for all the stars using the \textit{Gaia} photometry
together with 
an assumed distance of 18\,kpc, $A(V)=0.048$ and an assumed mass of $0.8\,\rm M_{\sun}$.
The method is iterative and was described in detail in 
\citet{lombardo21}. In short: The effective temperature was derived from
the $G_{BP}-G_{RP}$ colour by interpolating in a grid of synthetic colours,
then the surface gravity was derived from the Stefan-Boltzmann equation for the
given effective temperature, observed $G$, and distance. At each step
in the iteration, extinction coefficients were interpolated in a grid of synthetic
coefficients. The resulting dereddened colours, along with the \textit{Gaia} photometric errors,
are provided in Table \ref{table:idphotvr}. The effective temperatures
and surface gravities are provided in Table\,\ref{table:par}.
The error on the effective temperatures is dominated by the
photometric error on the $G_{BP}-G_{RP}$ colour. While for our stars, the \textit{Gaia} $G$ magnitudes
have a precision of a few milli-magnitudes, the colours have photometric errors
of up to 0.076 magnitudes, which translates into errors  in effective temperature 
of up to 300\,K. We interpret this as a combined effect of the narrower bands
and the fewer transits in $G_{BP}$ and $G_{RP}$ with respect to $G$.
Had we used the calibration of \citet{mucciarelli2021} to derive effective temperatures
we would have derived temperatures that are 71\,K cooler on average, with
a standard deviation of 42\,K. We consider this systematic error negligible
with respect to the errors implied by  the photometric errors.

The error in surface gravity is dominated by the distance estimate (see Sect.\,\ref{sec:distance} )
and amounts to 0.12\,dex from 18\,kpc to 20.9\,kpc.
The error on $G$ is negligible; the error on $G_{BP}-G_{RP}$, however,
implies an error on \logg\ in the range 0.04 to 0.08 \,dex. The errors on \logg\
provided in Table \ref{table:par} were obtained by adding these two errors
in quadrature.

Keeping the effective temperature and gravity fixed, we derived the chemical abundances
of Fe and Mg using \mygi\ \citep{mygi}. The grid of theoretical spectra used with
\mygi\ was computed from a grid of ATLAS 12 models (L. Sbordone, priv. comm.
\citealt{K05}), and the atomic data used were those of \citet{heiter21}.
The results are provided in Tables\,\ref{table:idphotvr} and \ref{table:mgca}.
 The systematic uncertainties on the abundances due to the uncertainties in the atmospheric parameters
were investigated by \citep{toposV} for star
SDSS\,J154746+242953, whose atmospheric parameters are similar to those of our stars.
Their investigation also used X-shooter spectra. For Mg, the effect of an error of 100\,K in \teff\ is 0.1\, dex,
and for an error of 0.3\,dex in \logg\, it is $-0.1$\,dex. A change of 0.5\,\kms\ results in a change of $-0.15$\,dex.

The 15 observed stars clearly show a sizeable spread in [Fe/H] that ranges from --3.4 to --1.7, which is at odds with the results obtained for the sample of giant stars in \citet{martin22}, \citet{pristineXVII}, and \citet{viswanathan2024}. This spread may be genuine, but may also be driven by the contaminants in the sample.
In order to ensure that this result is solid, one of us (PFR) reanalysed the sample
of stars using MARCS models and line-profile fitting using synthetic spectra
computed with {\tt turbospectrum} \citep{ap98,turbo}. This independent analysis
confirmed the results presented in Table~2.

Figure~\ref{vrad} shows the radial velocities of the stars in our sample. Overall, the sample overlaps in velocity with stream  members that were confirmed by \cite{martin22} and \cite{pristineXVII}. As mentioned above, however, the metallicities show a wider spread when taken at face value. One of the goals when we gathered this sample was to confirm the fairly large velocity dispersion of the C-19 stars and their uniform metallicities. This critically depends on distinguishing  genuine C-19 members from halo contaminants. All stars were selected from the STREAMFINDER sample of C-19 and therefore have proper motions, distances, and positions in the colour-magnitude diagram that are by design consistent with them being members of the C-19 stream. We can only rely on the radial velocities and metallicities obtained from the X-shooter spectra to assess membership.

We started by building a generic likelihood model that is agnostic about the membership or contaminant status of each observed star. This model was defined over the range $-350<V_r<-160$\,\kms\ and $-3.7<{\rm [Fe/H]}<-1.6$, where all observed stars are located. It included a C-19 component, modelled as a normal function in velocity and metallicity space, and a contamination component, also modelled as a two-dimensional normal function in both dimensions. The likelihood is therefore

\begin{dmath*}
\mathcal{L}\left(\{V_{r,i},\delta V_{r,i},{\rm [Fe/H]}_i,\delta{\rm [Fe/H]}_i\}|\eta,\langle v_r\rangle, \sigma_{vr},\langle {\rm [Fe/H]}\rangle,\sigma_{\rm [Fe/H]} \right) \\
= \prod_i\left\{(1-\eta) \mathcal{N}(V_{r,i},\delta V_{r,i}|\langle v_r\rangle,\sigma_{vr})\times\\
\mathcal{N}\left({\rm [Fe/H]}_i,\delta{\rm [Fe/H]}_i|\langle {\rm [Fe/H]}\rangle,\sigma_{\rm [Fe/H]}\right)\\
+\eta \mathcal{N}(V_{r,i},\delta V_{r,i}|\langle v_{r,\rm cont}\rangle,\sigma_{vr,{\rm cont}})\times\\
\mathcal{N}\left({\rm [Fe/H]}_i,\delta{\rm [Fe/H]}_i|\langle{\rm [Fe/H]}_{\rm cont}\rangle,\sigma_{\rm[Fe/H],cont}\right)\right\}.
\end{dmath*}

\noindent Here, $\mathcal{N}$ is the normal function defined as

\begin{dmath*}
\mathcal{N}(x,\delta x|\mu,\sigma) = \frac{A}{\sqrt{2\pi}\sigma_{\rm tot}}\exp\left(-\frac{1}{2}\left(\frac{x-\mu}{\sigma_{\rm tot}}\right)^2\right),
\end{dmath*}

\noindent with $\sigma_{\rm tot}^2 = \sigma^2+\delta x^2$ and $A$ a normalisation constant such that the normal function integrates to unity for the range over which it is defined. The set $\{V_{r,i},\delta V_{r,i},{\rm [Fe/H]}_i,\delta{\rm [Fe/H]}_i\}$ represents the set of X-shooter velocities and metallicities and the associated uncertainties, as listed in Tables~1 and~2. The parameters of interest are $\left(\eta,\langle v_r\rangle,\sigma_{vr},\langle {\rm [Fe/H]}\rangle,\sigma_{\rm [Fe/H]}\right)$, which correspond to the contamination fraction, the velocity mean and dispersion, and the metallicity mean and dispersion, respectively.

To simplify the calculations and because the STREAMFINDER selection of potential C-19 members makes it very likely that any contaminant belongs to the halo, we fixed the parameters of the contamination part of the model to generic expectations of the velocity and metallicity distribution of halo contaminants: $\langle v_{\rm r,cont}\rangle=-180$\,\kms (which is equivalent to $v_{\rm gsr}\sim0$\,\kms), $\sigma_{v_{r,\rm cont}}=100$\,\kms, $\langle{\rm [Fe/H]}_{\rm cont}\rangle=-1.5$, and $\sigma_{\rm[Fe/H],cont}=0.3$. Since the model was only defined over a limited velocity and metallicity range, the metallicity parameters, while simplistic, mainly aimed to reproduce the slope of the halo metallicity distribution function in the regime of interest, $\rm [Fe/H]<-1.6$.

\begin{figure}
\resizebox{\hsize}{!}{\includegraphics[clip=true]{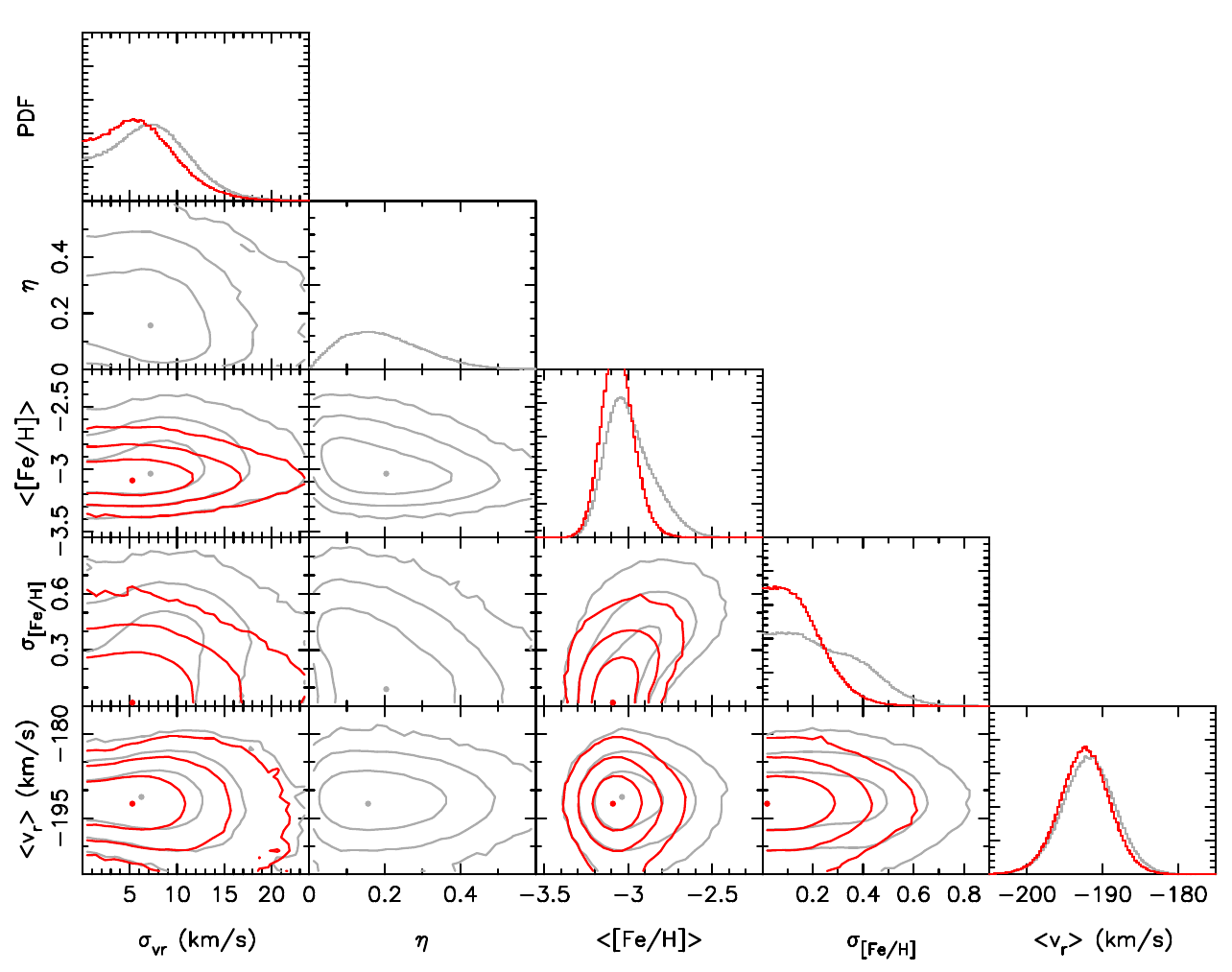}}\hfill
\caption{Probability distribution functions for the five parameters of the inference on the properties of the combined velocity and metallicity distribution of C-19 members. The five parameters are the fraction of contaminants, $\eta$, and the mean and dispersion of the velocity ($\langle v_r\rangle$ and $sigma_{vr}$) and metallicity ($\langle {\rm [Fe/H]}\rangle$ and $\sigma_{\rm [Fe/H]}$) parts of the model. The top panel of each column shows the one-dimensional marginalised probability distribution functions from which the favoured parameter models and related uncertainties are determined. The red contours and histograms show the same after the three likely contaminant stars are removed and the MW contamination is fixed to $\eta=0$.}
\label{corner}
\end{figure}

We explored the parameter space with our own implementation of a Metropolis-Hastings algorithm \citep{Metropolis1953,Hastings1970}, and we evaluated the combined posterior distribution on these five parameters assuming flat priors. The marginalised one- and two-dimensional probability distribution functions (PDFs) are shown in grey in Fig.~\ref{corner}. The model is ambivalent about the overall properties of C-19 based on the X-shooter sample, and this is reflected in particular in the poor constraints on the metallicity parameters. We note, however, that the model favours a non-negligible fraction of contaminants, which in most cases translates into an expectation of two to three contaminants in the sample of 15 stars observed with X-shooter. In particular, stars S05 (high-velocity offset and high metallicity), S12 (somewhat offset in velocity and high metallicity), and S13 (same diagnostic as S12) have high probabilities (>0.7) of being contaminant members for most evaluations of the likelihood in the MCMC chain. S06 is the  only other star that is not unambiguously a C-19 member, but whose probability of being a contaminant is significantly lower (<0.3; somewhat offset in velocity and fairly high metallicity, but with a large uncertainty).

\begin{figure}
\centering
\resizebox{7.7 truecm}{!}{\includegraphics[clip=true]{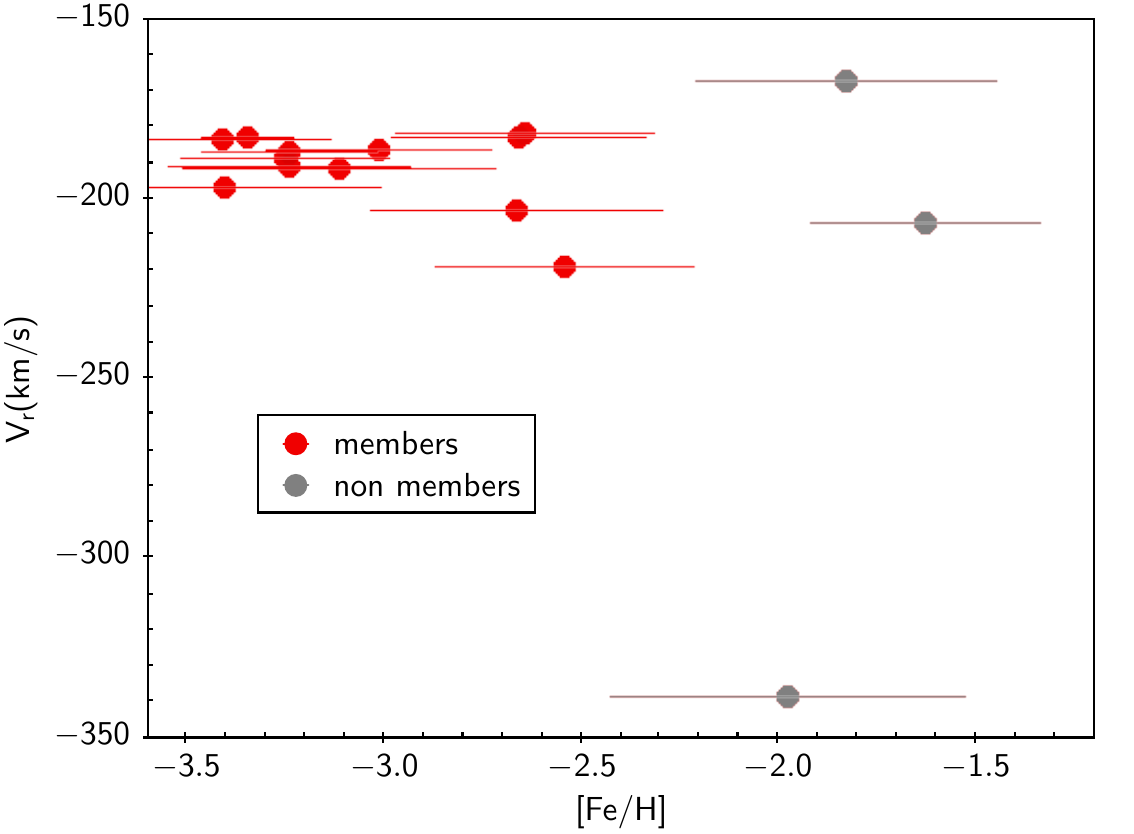}}
\caption{[Fe/H] vs. radial velocity for all the observed stars. The likely member stars are shown in red,
and the three likely contaminants are shown in grey.}
\label{feh_vrad}
\end{figure}

To better visualise the situation, we show in Fig.\,\ref{feh_vrad}  [Fe/H] versus radial velocity for all the 15 stars
observed with X-shooter.
Inspired by these results, we also inferred the properties of the C-19 members after removing the three likely contaminants (S05, S12, and S13) and for a model without contamination ($\eta=0$). The resulting posterior PDFs are displayed in red in Fig.~\ref{corner} and show properties that are better constrained, but also PDFs that are better behaved. In what follows, we base our discussion on these posterior PDFs, but we recognise that they are sensitive to our choice of removing the three likely contaminants.

The inference on the mean radial velocity of the 
X-shooter sample, $\langle v_r\rangle=-192\pm3$\,\kms, agrees well with previous studies. 
The constraint on the velocity dispersion of the sample, $\sigma_{vr}=5.9^{+3.6}_{-5.9}$\,\kms, 
is unfortunately,plagued by the large systematic uncertainty floor 
that stems from the X-shooter spectrograph. 
While it perfectly agrees with the previous estimate 
by \citet[][$\sigma_{vr}=6.2^{+2.0}_{-1.4}$\,\kms]{pristineXVII}, 
it does not lead to a better-constrained parameter.

For the metallicity part of the model, we inferred a 
mean metallicity $\langle{\rm [Fe/H]}\rangle=-3.1\pm0.1$ and a dispersion that 
formally favours a non-zero dispersion, but is 
entirely consistent with zero: $\sigma_{\rm [Fe/H]}=0.09_{-0.09}^{+0.13}$, or $\sigma_{\rm [Fe/H]}<0.35$ 
at the 95\% confidence level, 
to be compared with $\sigma_{\rm [Fe/H]}<0.18$ at the 95\% confidence 
level as determined by \citet{martin22} from a smaller sample of brighter stars.

Overall, we confirm previous results from \citet{martin22}, \citet{pristineXVII}, and \citet{viswanathan2024}, 
but the new velocities and metallicities, taken independently from previous studies, 
do not lead to stronger constraints on the velocity and metallicity dispersions of C-19 stars.

\section{Colour-magnitude diagram, distance, and age\label{sec:distance}}
\begin{figure*}
\resizebox{\hsize}{!}{\includegraphics[clip=true]{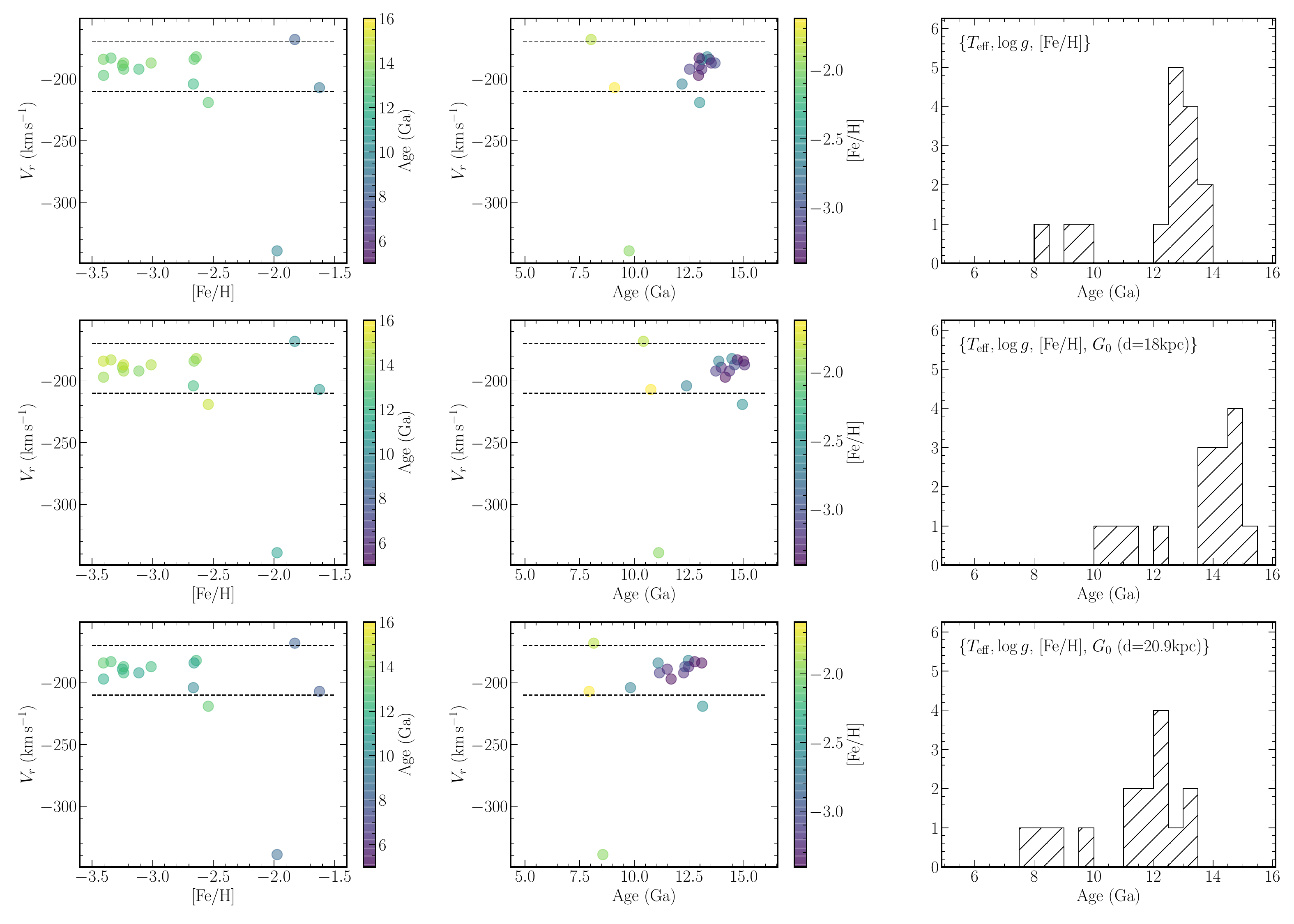}}
\caption{ Radial velocities as a function of metalicity and age and age histogram.
Left: 
Radial velocity vs. [Fe/H] colour-coded by age. 
Middle: Radial velocity vs. age colour-coded by [Fe/H]. 
Right: Age histogram. 
Each row shows results obtained with the projection of different parameters, 
as indicated in the top left corner of the histogram panels.  }
\label{ages_vr}
\end{figure*}

\begin{figure}
\resizebox{\hsize}{!}{\includegraphics[clip=true]{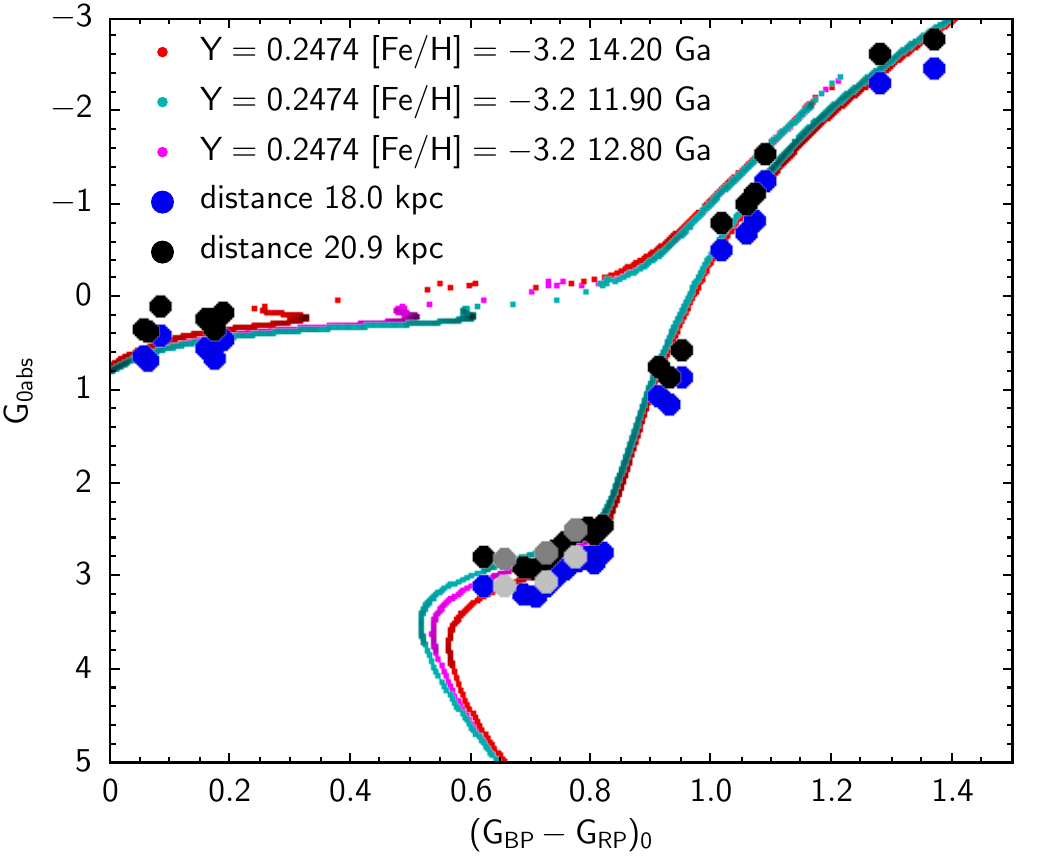}}
\caption{\textit{Gaia} colour-magnitude diagram of all the C-19 members studied
spectroscopically. The absolute magnitudes are shown for two
possible distances: 18\,kpc (blue), and 20.9\,kpc (black).
The three  non-member stars are highlighted in light grey (18\,kpc)
and grey (20.9\,kpc).
For comparison, three BASTI $\alpha-$enhanced isochrones \citep{pietrinferni21}
 of [Fe/H]=--3.2 of 14.20\,Ga (red), 12.80\,Ga (cyan), and 11.90\,Ga (magenta) are shown.}
\label{cmd}
\end{figure}

One of the goals of the observations was to constrain the age of C-19 from
a study of the colour-magnitude diagram of its confirmed members.
For this purpose,
we adopted the pipeline of \citet{Kordopatis23a} to obtain an age estimation for the stars via the projection of different sets of parameters on $\alpha-$enhanced BaSTI isochrones ([$\alpha$/Fe]=0.4, Y=0.247) up to $\tau=16$\,Ga. In Table~\ref{tab:gk_ages}, the results of three different projections are considered, which were obtained using
\begin{itemize}
\item 
\teff, \logg\ and [FeI/H] (labelled $\tau_1$);
\item 
\teff, \logg,  [FeI/H] and absolute $G_0$ assuming d=18\,kpc  (labelled $\tau_2$);
\item 
\teff, \logg -0.12,  [FeI/H] and absolute $G_0$ assuming d=20.9\,kpc  (labelled $\tau_3$).
\end{itemize}
For $\tau_3$, the surface gravities are corrected by $-0.12$ to account for the larger adopted distance.

The ages as a function of $Vr$ and [Fe/H] are shown in Fig.~\ref{ages_vr}. Whereas the age values may change between the different projections, values consistently older than 11\,Gyr are found for the bulk of the stars.
Depending on the projection, three to five stars stand outside the age-Vr clump. 
Amongst these,  stars S12 and S13 are always too young, and S04 appears to be too young
in the $\tau_2$ and $\tau_3$ projections.
Quite interestingly, star S04 is also at the edge of the metallicity distribution,
although it has almost the same metallicity as S10. 
It would therefore not be possible to exclude it as member simply because of its metallicity.

In Fig.\,\ref{cmd} we show the \textit{Gaia} colour-magnitude diagram for all
the C-19 members that were studied spectroscopically.
The \textit{Gaia} broad-band colours were corrected assuming the
reddening from the maps of \citet{Schlafly}, and the extinction coefficients
were interpolated in  a grid of theoretical values (Mucciarelli et al. in preparation)
during the process of parameter determination. 
As a comparison, we used the most metal-poor $\alpha-$enhanced BASTI
isochrones \citep{pietrinferni21} available at [Fe/H]=--3.2.
We chose the helium abundance Y=0.247 because the use of  higher available He abundances
had second-order effects on the subgiant and the red giant branch.
We show three isochrones corresponding to the average age of the bulk of the subgiant (SG) stars
in the different  projections: 12.80 Ga ($\tau_1$), 14.23 Ga ($\tau_2$), and 11.90 Ga ($\tau_3$). 
For the preferred distance of 18\,kpc, 
the implied age is  14.1\, Ga, which is older than the age of the Universe
\citep[$13.7^{+0.3}_{-0.2}$ Ga][]{beyondplanckXII}, 
as derived from the fluctuations of the 
cosmic microwave background, measured by the Planck mission \citep{PlanckI}.
When we instead consider the distance 20.9\,kpc, which is preferred from the
\textit{Gaia} parallaxes \citep{martin22} and measured photometric distances using Bayesian inference on photometry, parallax, and metallicity \citep{viswanathan2024}, 
the implied age is 13.43\,Ga, which is slightly younger than the age
of the Universe.

\section{Discussion}

\begin{figure}
\resizebox{\hsize}{!}{\includegraphics[clip=true]{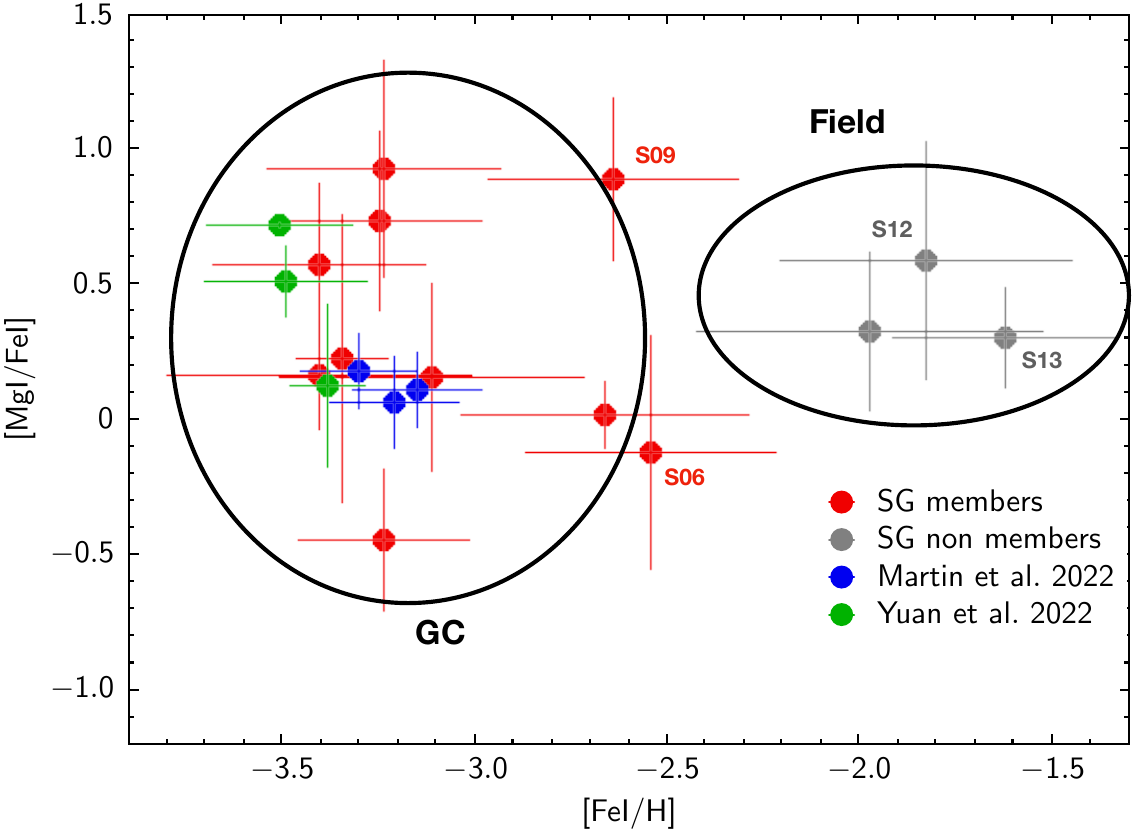}}
\caption{[Fe/H]--[Mg/Fe] diagram for our programme stars and the stars from 
\citet{martin22} and \citet{pristineXVII} for which Mg abundances are available.}
\label{mgfe}
\end{figure}

\begin{figure*}
\centering
\includegraphics[width=0.9\textwidth]{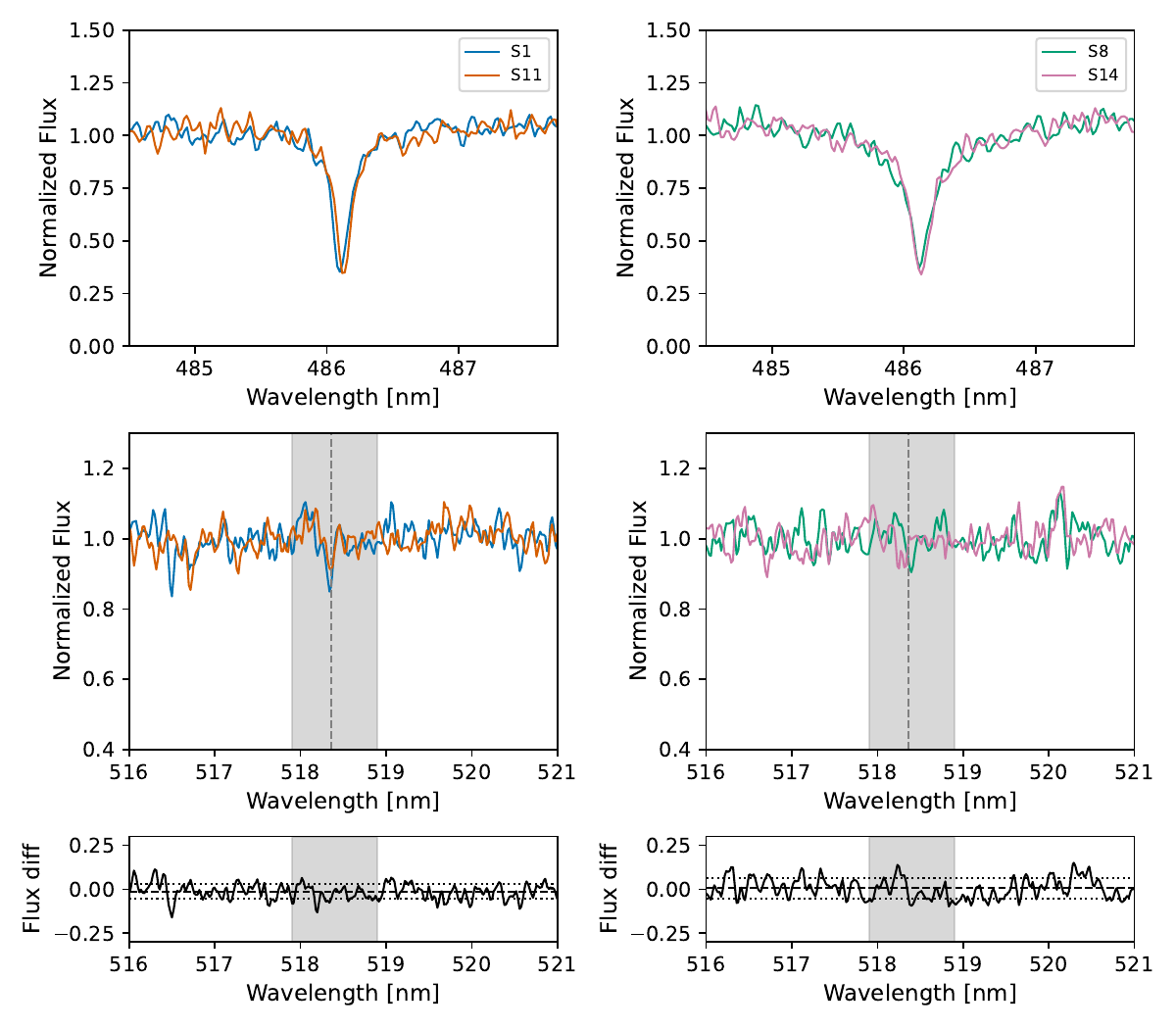}
\caption{Comparison of spectra for pairs of stars with similar atmospheric parameters (S1/S11 and S8/S14). The top panel displays a zoomed view around the strong H$\beta$ line, revealing virtually identical profiles. The middle panel shows the spectra around the Mg Ib 518.3 nm line. The bottom panel shows the difference between the two spectra.
Given the near-identical temperatures, gravities, and metallicities within each pair, the differences in the strengths of 
the \ion{Mg}{i}b 518.3 line are solely attributed to intrinsic variations in the Mg abundances.}
\label{mgspectra}
\end{figure*}

\begin{figure}
\resizebox{\hsize}{!}{\includegraphics[clip=true]{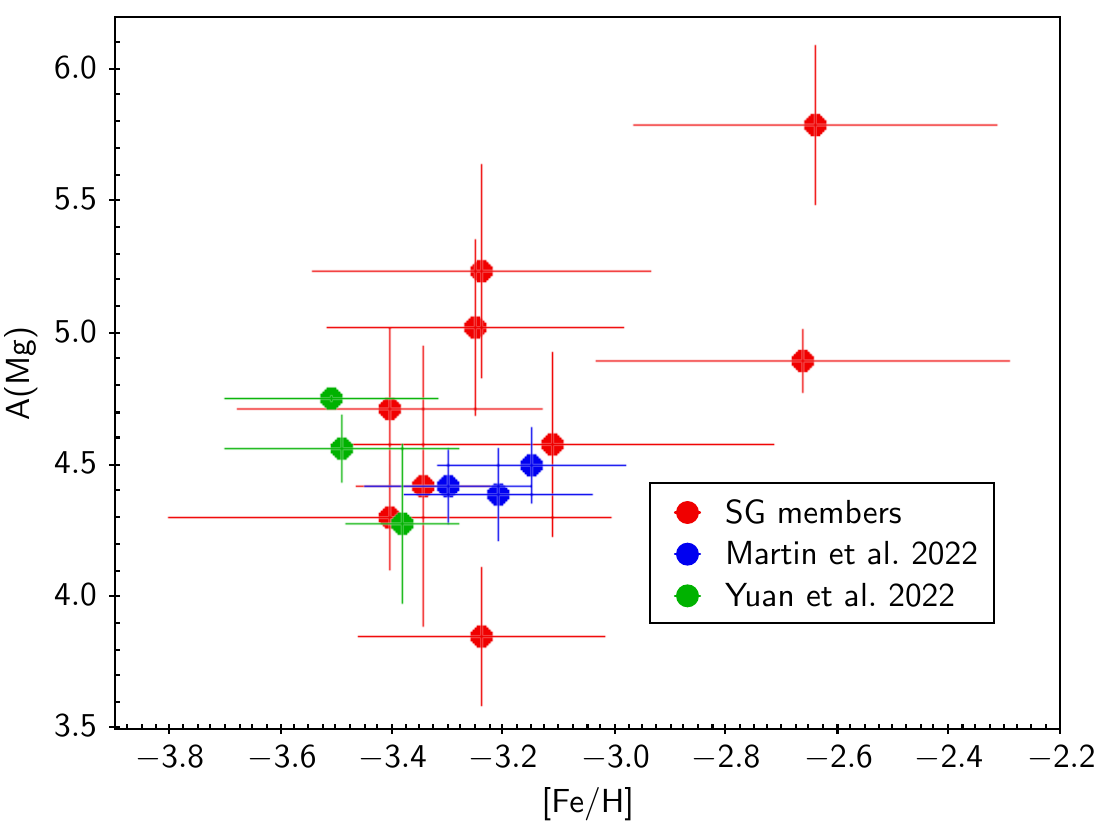}}
\caption{[Fe/H]--A(Mg) diagram for our  stars that are likely members of C-19
and the stars from 
\citet{martin22} and \citet{pristineXVII} for which Mg abundances are available.}
\label{amg}
\end{figure}

\begin{figure}
\centering
\resizebox{\hsize}{!}{\includegraphics{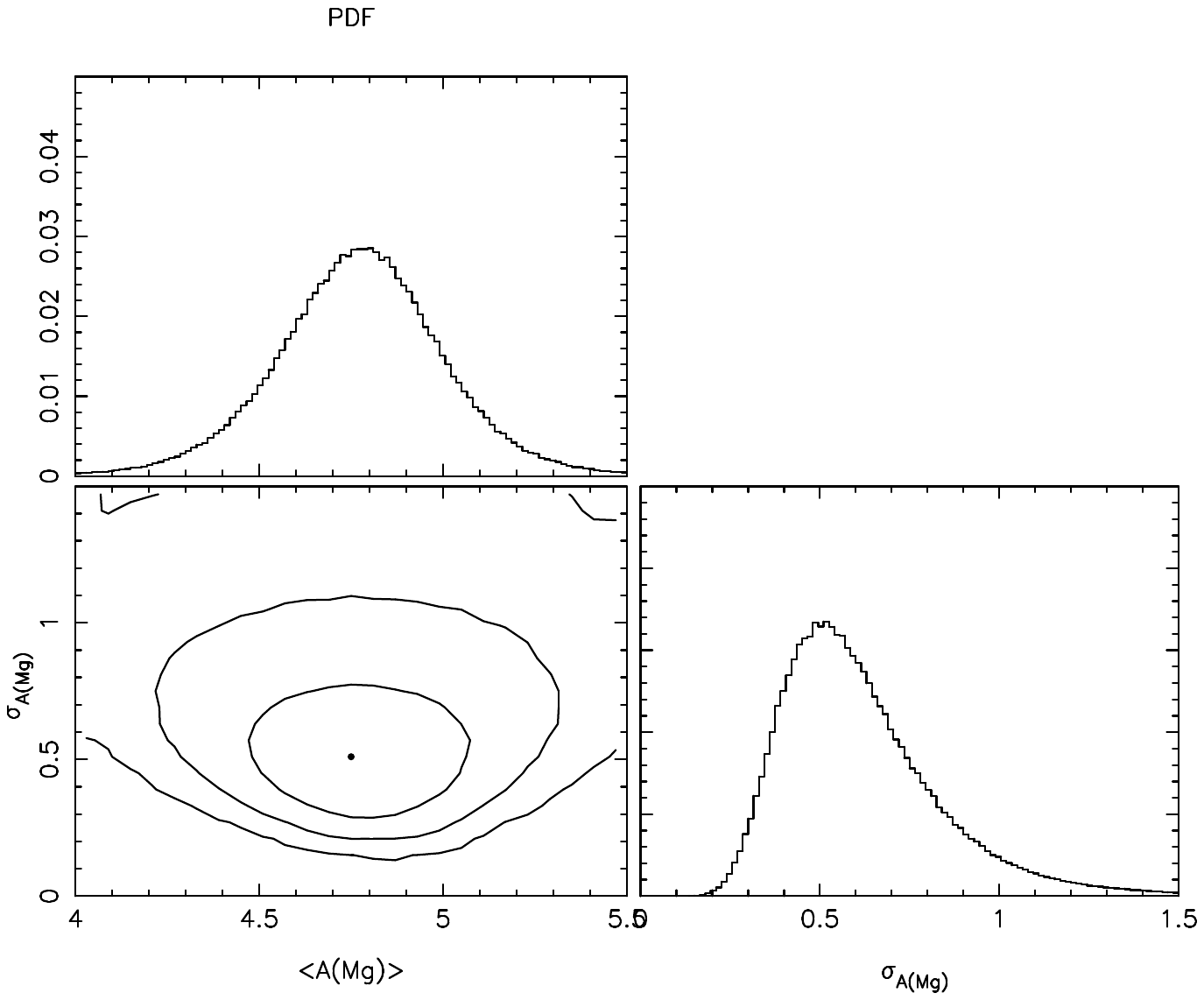}}
\caption{Probability distribution function for the nine stars identified
as C-19 members and a measured Mg abundance, excluding S06, whose membership is questionable.}
\label{mgpdf}
\end{figure}

\begin{figure}
\centering
\resizebox{\hsize}{!}{\includegraphics{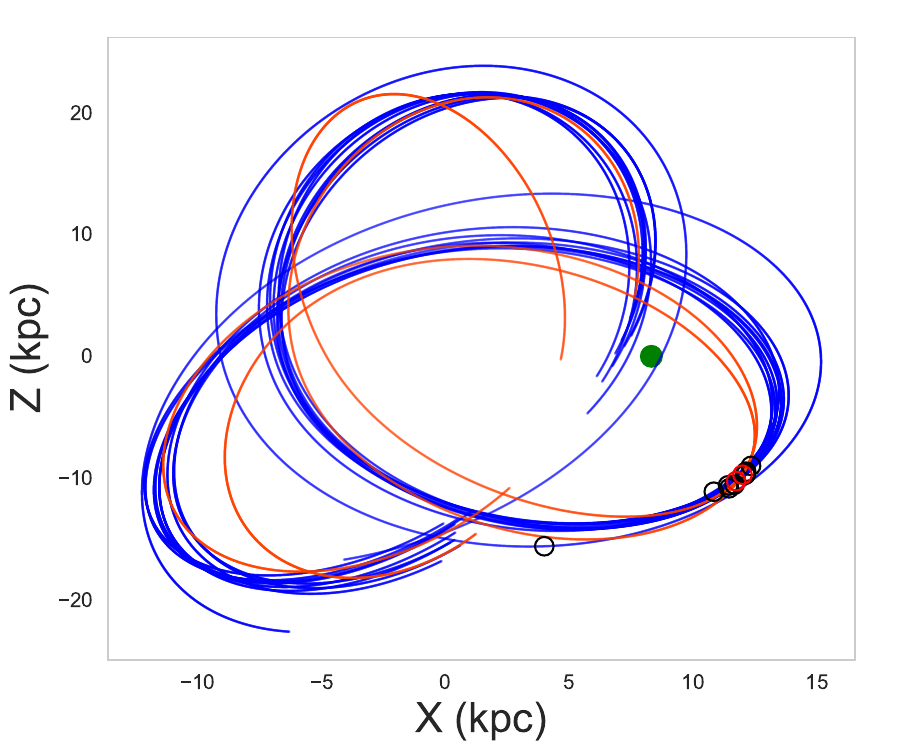}}
\caption{Orbits of the C-19 in \citet{pristineXVII} (black circles and blue orbits) and stars S12 and S13
(orange) in galactocentric
coordinates (the Sun is shown by the green circle).}
\label{orbit}
\end{figure}

Our statistical analysis of the 15 stars in the X-shooter sample suggests that most of the observed metallicity spread is driven by the three  halo contaminants (stars S05, S12, and S13). The resulting velocity and metallicity properties of the sample of 12 likely members yield results that are entirely consistent with those provided for C-19 by \citet{martin22} and \citet{pristineXVII}. In particular, we derive $\sigma_{vr}=5.9^{+3.6}_{-5.9}$\,\kms and $\sigma_{\rm [Fe/H]}<0.35$ at the 95\% confidence level.

We integrated the orbits for the 12 radial velocity members.
The orbits of stars S12 and S13, regardless of the assumed
distance, are clearly different from those of the other
ten radial velocity members, which instead remain
very close to each other. This is further evidence that
S12 and S13 may indeed not belong to C-19.
We conclude that the data support
a single metallicity for the C-19 stream, with an unresolved metallicity
dispersion. 
The analysis of the ages in Sect.\ref{sec:distance} suggests that  S04 may also be an interloper.

Further clues come from the [Fe/H]--[Mg/Fe] diagram
shown in Fig.\,\ref{mgfe}. For all the C-19 member stars, including those in
\citet{martin22} and \citet{pristineXVII}, we plot [\ion{Mg}{i}/\ion{Fe}{i}] versus
[FeI/H] LTE abundances in order to be on the same scale as that of our SG
stars. The use of NLTE corrections on [\ion{Mg}{i}/\ion{Fe}{ii}] and [\ion{Fe}{ii}/H], when available, 
does not change the general picture. The majority of the
stars of C-19 that were subject to a chemical analysis clearly belong to
the metal-poor population (13 out of 16). Moreover, the metal-poor component
has a sizeable dispersion in [Mg/Fe].
 To better visualise the Mg abundances, we plot Fig.\,\ref{amg}
A(Mg) only for the member subgiant stars and for the giant stars in \citet{martin22} and \citet{pristineXVII}.
When the whole sample is considered together (15 stars, nine SG, and six giants )
the dispersion in A(Mg) is 0.44 dex, and the mean uncertainty on Mg is 0.25\,dex.
We interpret this as  evidence that
the dispersion in the Mg abundances among C-19 stars is higher than what can be expected
based on observational uncertainties alone.

To further substantiate the evidence of intrinsic variation in Mg among the C-19 SG stars, we selected two pairs of stars with similar atmospheric parameters (S1/S11 and S8/S14; see Table~\ref{table:par}). The top panel of Fig.~\ref{mgspectra} displays a zoomed view around the strong H$\beta$ line for these stars and reveals virtually identical profiles. The middle panel shows the spectra of the selected stars around the \ion{Mg}{I} 518.3 nm line. Stars S1 and S11, as well as stars S8 and S14, exhibit virtually identical stellar parameters as observed, but differences may be noted in the strength of the Mg line. This further confirms the evidence for an intrinsic Mg spread  in C-19, as  already seen among giant stars.

Finally, we considered the probability distribution function of the Mg abundance for 
the nine C-19 members with a measured Mg, but we excluded S06, whose
membership is questionable.
In Fig.\,\ref{mgpdf} we show the relevant corner plot. It clearly shows that a dispersion in Mg
abundances is favoured. The result would be the same had we included S06. The dispersion in Mg is
strongly supported by the dispersion found among giants \citep{martin22,pristineXVII}, and the data
of the subgiants are consistent with this dispersion.
Statistically, the additional dispersion is only detected at 1.3$\sigma$ when we consider the subgiant sample alone.
When we arbitrarily remove the two stars with the larger error
on the Mg abundance, the dispersion on A(Mg) decreases to 0.56\,dex and the mean error
drops to 0.27\,dex. This means that the detection would be at 2$\sigma$.

The Mg-Al anti-correlation, typical of GCs \citep[see e.g.][]{gratton01}, is not always observed. In particular, it
is not observed in the less massive clusters  or in the most metal-rich clusters \citep[e.g.][]{pancino2017}.
The significant spread in the Mg abundances may suggest that the progenitor
of C-19 was as massive as NGC\,4833, which displays a well-developed Mg-Al anti-correlation
\citep{pancino2017} and has a present-day mass
of about $1.6\times 10^5~ M_\odot$ \citep{mandushev1991}. 
For NGC\,4833, the database called Fundamental parameters of 
Galactic globular 
clusters\footnote{\url{https://people.smp.uq.edu.au/HolgerBaumgardt/globular/parameter.html}}
reports an initial mass of $10^6~M_\odot$ .
The present-day and the initial  mass of NGC\,4833 
are much higher than the minimum mass of C-19 of $0.8\times 10^4~M_\odot$
determined by \citet{martin22}. 
It is noteworthy that the spread of Mg among the C-19 stars appears to be larger than that measured in
the nearby halo stellar stream  ED-2 \citep{Dodd2023},  which was suggested 
to have originated from a disrupted low-mass star cluster \citep{Ceccarelli2024,Balbinot2024}. 
In their study, \citet{Balbinot2024} inferred a plausible mass range for the cluster, which appears to be relatively narrow, between $2\times10^3M_\odot$ and $5.2\times10^4M_\odot$, similar to the current stellar mass of  C-19.
With a measured metallicity of [Fe/H]$_{\rm II} =-2.46\pm0.02$, ED-2 is significantly richer in metals than C-19. The evidence for a prominent Mg-Al correlation in C-19, compared to the small scatter observed in Mg and Al in ED-2 \citep{Ceccarelli2024,Balbinot2024}, agrees with the notion that in clusters (or progenitors) with comparable masses, the amplitude of the Mg-Al anti-correlation is larger for more metal-poor systems.

Two of the stars in \citet{pristineXVII} have measured Al abundances. As discussed
in that paper, they support the existence of an Mg-Al anti-correlation.
All the measured Na abundances also support a dispersion in Na, as expected in
a GC showing an Na-O anti-correlation. 
Unfortunately, the quality of the X-shooter spectra does not allow us to measure Na
in any of the SG stars. If it could be shown that the metal-poor
component includes a dispersion in Na abundances, that would be a strong
indication that the metal-poor component is indeed a GC.
Arguing for a GC origin on the sole basis of a negligible metallicity dispersion is inconclusive because we know of ultra-faint dwarf galaxies with very small metallicity dispersions \citep[see e.g.][]{vargas13}.

Concerning the radial velocities, 
\citet{pristineXVII} argued that
the velocity dispersion of C-19 is significantly higher than that of streams that are
firmly associated with disrupted GCs.
Along the same lines, \citet{pristineXVIII} have argued that the width and
velocity dispersion of C-19 are larger than what can be obtained by disrupting
a GC, and they are similar to what is expected from disrupting dwarf galaxies.
Since C-19 approaches a turning point in the orbit (see Fig. \ref{orbit}),  
its line-of-sight velocity dispersion may increase \citep[e.g.][]{helmiwhite1999, viswanathan2024}, but the $N$-body simulations studied in \citet{pristineXVIII} suggest that on the specific orbit of C-19, this increase alone is not sufficient to explain the radial velocity dispersion of C-19 using a simple GC progenitor.
Given the lower precision of our radial velocity measurements, we can only confirm the previously determined velocity dispersion. As mentioned above, the large dispersion in Mg abundances does suggest a high mass
for the C-19 progenitor. 

Our favourite interpretation still is that C-19 is a disrupted GC. This view is mainly supported by the spread in [Mg/Fe] abundances and, less strongly, also by
the unresolved metallicity dispersion. 
We cannot completely rule out the possibility that C-19
is a disrupted galaxy, however. In this case, it should have contained a GC
(to explain the dispersion in [Mg/Fe]), which would account
for most of the confirmed C-19 members. The  three higher-metallicity
stars reported in this study and highlighted in Fig.\,\ref{mgfe} could then
be interpreted as field stars 
of the parent galaxy. This scenario is also supported by the combined findings of \cite{Malhan_2021, Malhan_2022}, who suggested that the C-19 stream was accreted inside a (now disrupted) high-redshift dwarf galaxy that gave rise
to the LMS-1/Wukong merger \citep{Naidu2020}. The stars of this
galaxy possessed a wide metallicity spread and reached higher values. This association
may be questioned because although both streams have polar orbits, their orbital poles point towards different directions, as shown by comparing Fig. 11 in this work and Fig. 4 in  \citet{yuan2020}.

Of the Local Group dwarf spheroidal galaxies, only Fornax \citep{Wilson1955}
and Sgr \citep{Ibata1994} host GCs. In Fornax, all the GCs are significantly
poorer in metals than the field stars \citep[see e.g.][and references therein]{Larsen2012}.
Sagittarius  has
the GC M54 at its centre, which is about 1\,dex poorer in metals than the
field stars among which it is embedded \citep{Mucciarelli2017,Minelli2023}.
The metal-poor field population of Sgr \citep{Bonifacio2006,Hansen2018,Sestito2024} is a minor component.
In a scenario of inside-out galaxy formation,
the field stars could then have a higher metallicity than the stars in a GC, having formed from gas already polluted by the supernovae
that exploded in the GC. If instead the galaxy formed outside-in, a metal-poor GC
may have formed in its halo and then spiralled to its centre due to dynamical friction.
If the GC were the centre of  the putative dwarf, 
then this would not help us to explain the width and dispersion of the stream. 
They would be more consistent with a GC that was disrupted 
in the gravitational potential of the dwarf before accretion, which implies that the GC was not at the centre
of the dwarf.
Although these scenarios may in principle
account for a  
mass of the C-19 progenitor that is higher than what can be accounted for by a GC, the current sample
does not support the notion that the field population can account for a mass
higher than half of the GC. This is probably not sufficient
to explain the observed dispersion in radial velocity. 
In any case, even these scenarios require that a GC can be formed at this extremely
low metallicity. 

\begin{acknowledgements}
We are grateful to M. Bellazzini for his careful reading and critical assessment
of our manuscript.
PB is supported by the European Research Council through grant 835087 (SPIAKID).
We gratefully acknowledge support from the French National Research Agency (ANR) funded project 
``Pristine'' (ANR-18-CE31-0017).
JIGH acknowledges financial support from the Spanish Ministry of Science, Innovation and Universities (MICIU) project PID2020-117493GB-I00. 
GT acknowledges support from the Agencia Estatal de Investigaci{\'o}n del Ministerio de Ciencia en Innovaci{\'o}n (AEI-MICIN) and the European Regional Development Fund (ERDF) under grant number PID2020-118778GB-I00/10.13039/501100011033 and the AEI under grant number CEX2019-000920-S.
G.K. and F.G. gratefully acknowledge support from the French National Research Agency (ANR) funded project
“MWDisc” (ANR-20-CE31-0004). E.S. acknowledges funding through VIDI grant "Pushing Galactic Archaeology to its limits" (with project number VI.Vidi.193.093) which is funded by the Dutch Research Council (NWO). This research has been partially funded from a Spinoza award by NWO (SPI 78-411). This research was supported by the International Space Science Institute (ISSI) in Bern, through ISSI International Team project 540 (The Early Milky Way).
RE acknowledges support from the National Science Foundation (NSF) grants AST-2206046 and AST-1909584.

This work has made use of data from the European Space Agency (ESA) mission
{\it Gaia} (\url{https://www.cosmos.esa.int/gaia}), processed by the {\it Gaia}
Data Processing and Analysis Consortium (DPAC,
\url{https://www.cosmos.esa.int/web/gaia/dpac/consortium}). Funding for the DPAC
has been provided by national institutions, in particular the institutions
participating in the {\it Gaia} Multilateral Agreement.
This research has made use of the SIMBAD database, operated at CDS, Strasbourg, France.

\end{acknowledgements}

   \bibliographystyle{aa} 

   \bibliography{ref_c19} 

\begin{appendix}
\onecolumn
\section{Tables}

This appendix contains the tables with all the data cited in the paper.

\begin{table}[H]
\caption{Identification and photometry of our program stars.}
\label{table:idphotvr}
\centering
\begin{tabular}{lrrcccccc}
\hline
  \multicolumn{1}{c}{Star} &
  \multicolumn{1}{c}{GaiaDR3sourceid} &
  \multicolumn{1}{c}{$G_0$} &
  \multicolumn{1}{c}{$\delta(G)$} &
  \multicolumn{1}{c}{$(G_{BP}-G_{RP})_0$} &
  \multicolumn{1}{c}{$\delta(G_{BP}-G_{RP})$} &
  \multicolumn{1}{c}{$V_r$} &
  \multicolumn{1}{c}{$\delta(V_r)$} &
  \multicolumn{1}{c}{Member} 
\\
 & & mag & mag& mag & mag & $\rm kms^{-1}$ & $\rm kms^{-1}$\\
\hline
  S01 & 2826300927030372224 &  19.246& 0.003   & 0.878& 0.043   &$-189$&   9 & yes\\   
  S02 & 2826352947674483840 &  19.371& 0.002   & 0.827& 0.056   &$-197$&   9 & yes\\   
  S03 & 2826454824298642560 &  19.196& 0.002   & 0.897& 0.041   &$-192$&   9 & yes\\   
  S04 & 2826456439206381312 &  19.534& 0.003   & 0.693& 0.055   &$-204$&   9 & yes\\   
  S05 & 2827756203686469120 &  19.548& 0.003   & 0.726& 0.076   &$-339$&   9 & no\\    
  S06 & 2827731670833206272 &  19.666& 0.004   & 0.825& 0.069   &$-219$&   9 & yes(?)\\   
  S07 & 2864770364985943424 &  19.252& 0.002   & 0.843& 0.049   &$-192$&   9 & yes\\   
  S08 & 2827700678349232128 &  19.624& 0.003   & 0.759& 0.069   &$-184$&   9 & yes\\   
  S09 & 2828240980938817792 &  19.744& 0.003   & 0.740& 0.062   &$-182$&  15 & yes\\   
  S10 & 2864329185945227904 &  19.442& 0.003   & 0.801& 0.046   &$-184$&   9 & yes\\   
  S11 & 2852259503209614336 &  19.267& 0.002   & 0.865& 0.041   &$-187$&   9 & yes\\   
  S12 & 2865245186507418496 &  19.338& 0.003   & 0.896& 0.055   &$-168$&   8 & no\\    
  S13 & 2864431371806333440 &  19.443& 0.002   & 0.774& 0.054   &$-207$&   8 & no\\    
  S14 & 2826664453062267776 &  19.510& 0.003   & 0.780& 0.075   &$-183$&   9 & yes\\   
  S15 & 2852422883765640320 &  19.606& 0.003   & 0.742& 0.053   &$-187$&   9 & yes\\   
\hline\end{tabular}       
\end{table}

\begin{table}[H]
\caption{Stellar parameters for our program stars.}
\label{table:par}
\centering
\begin{tabular}{lrrccccrcccc}
\hline
  \multicolumn{1}{c}{Star} &
  \multicolumn{1}{c}{$\rm T_{eff}$} &
  \multicolumn{1}{c}{$\delta(\rm T_{eff})$} &
  \multicolumn{1}{c}{log g} &
  \multicolumn{1}{c}{$\delta($log g$)$} &
  \multicolumn{1}{c}{[Fe I/H]} &
  \multicolumn{1}{c}{$\delta$(Fe I)} &
  \multicolumn{1}{c}{N(Fe I)} &
  \multicolumn{1}{c}{[Fe II/H]} &
  \multicolumn{1}{c}{$\delta$(Fe II)} &
  \multicolumn{1}{c}{N(Fe II)} &
 \\
  & K &  K & [c.g.s] & [c.g.s] & dex & dex& &  dex & dex&  &\\

\hline
  S01 & 5597 & 152& 3.49&  0.13&$-3.25$& 0.27  & 6 &         &     & 0  \\
  S02 & 5726 & 207& 3.59&  0.13&$-3.40$& 0.40  & 4 & $-2.71$ & 0.30 & 1 \\
  S03 & 5507 & 137& 3.45&  0.13&$-3.24$& 0.30  & 13& $-3.06$ & 0.30 & 1 \\
  S04 & 6244 & 232& 3.81&  0.13&$-2.66$& 0.37  & 6 &         &     & 0  \\
  S05 & 6104 & 313& 3.78&  0.14&$-1.97$& 0.45  & 27& $-2.13$ & 0.38& 2  \\
  S06 & 5738 & 260& 3.71&  0.14&$-2.54$& 0.33  & 8 &         &     & 0  \\
  S07 & 5652 & 175& 3.52&  0.13&$-3.11$& 0.40  & 7 &         &     & 0  \\
  S08 & 5916 & 269& 3.76&  0.14&$-3.40$& 0.27  & 4 &         &     & 0  \\
  S09 & 6035 & 253& 3.84&  0.14&$-2.64$& 0.33  & 10&         &     & 0  \\
  S10 & 5794 & 173& 3.65&  0.13&$-2.66$& 0.32  & 9 &         &     & 0  \\
  S11 & 5559 & 140& 3.50&  0.13&$-3.24$& 0.22  & 10&         &     & 0  \\
  S12 & 5667 & 201& 3.52&  0.13&$-1.83$& 0.38  & 6 &         &     & 0  \\
  S13 & 5837 & 205& 3.67&  0.13&$-1.62$& 0.29  & 56& $-1.61$ & 0.32& 4  \\
  S14 & 5840 & 289& 3.69&  0.15&$-3.34$& 0.12  & 4 &         &     & 0  \\
  S15 & 5984 & 209& 3.78&  0.13&$-3.01$& 0.28  & 7 &         &     & 0   \\
\hline\end{tabular}       
\end{table}

\begin{table}[H]
\caption{Magnesium and calcium abundance for our program stars.}
\label{table:mgca}
\centering
\begin{tabular}{lccrrrrr}
\hline
  \multicolumn{1}{c}{Star} &
  \multicolumn{1}{c}{A(Mg)} &
  \multicolumn{1}{c}{$\delta(Mg)$} &
  \multicolumn{1}{c}{[Mg/Fe]} &
  \multicolumn{1}{c}{N(Mg)} &
  \multicolumn{1}{c}{A(Ca)} &
  \multicolumn{1}{c}{[Ca/Fe]} &
  \multicolumn{1}{c}{N(Ca)} 
  \\
   & dex & dex & dex & & dex & dex& \\
\hline
  S01 & 5.02 &0.33 & $ 0.73$  & 3 &3.32 &$ 0.24$&  1 \\
  S02 & 4.30 &0.20 & $ 0.17$  & 3 &3.11 &$ 0.18$&  3 \\        
  S03 & 5.23 &0.40 & $ 0.93$  & 3 &3.30 &$ 0.21$&  1 \\ 
  S04 & 4.89 &0.12 & $ 0.02$  & 2 &3.27 &$-0.39$&  2 \\
  S05 & 5.89 &0.30 & $ 0.32$  & 4 &     &       &  0 \\ 
  S06 & 4.88 &0.43 & $-0.12$  & 3 &     &       &  0 \\     
  S07 & 4.58 &0.35 & $ 0.15$  & 4 &3.18 &$-0.04$&  3 \\     
  S08 & 4.71 &0.30 & $ 0.57$  & 1 &2.97 &$ 0.05$&  3 \\     
  S09 & 5.79 &0.30 & $ 0.89$  & 1 &3.10 &$-0.59$&  2 \\     
  S10 &      &     &          & 0 &3.30 &$-0.37$&  3 \\     
  S11 & 3.85 &0.26 & $-0.45$  & 3 &3.21 &$ 0.12$&  2 \\     
  S12 & 6.30 &0.44 & $ 0.59$  & 3 &4.64 &$ 0.14$&  3 \\     
  S13 & 6.22 &0.19 & $ 0.30$  & 6 &5.09 &$ 0.39$&  2 \\     
  S14 & 4.42 &0.53 & $ 0.22$  & 3 &3.05 &$ 0.06$&  2 \\     
  S15 &      &     &          & 0 &3.26 &$-0.06$&  2  \\    
\hline\end{tabular}\end{table}

\begin{table}[H]
    \centering
        \caption{Age determinations (in Ga) via isochrone projection of different parameters and their uncertainties.}
    \begin{tabular}{ccccccc}
    \hline
    Star & $\tau_1$ &  $\sigma(\tau_1)$& $\tau_2$ &  $\sigma(\tau_2)$& $\tau_3$ &  $\sigma(\tau_3)$ \\ \hline

S01  &  13.0  &  2.1  & 14.0  &  1.0  & 11.5  &  1.5 \\
S02  &  12.9  &  2.2  & 14.2  &  1.0  & 11.7  &  1.2 \\
S03  &  13.1  &  2.1  & 14.3  &  1.1  & 12.3  &  1.6 \\
S04  &  12.2  &  2.4  & 12.4  &  1.1  & 9.8  &  0.8 \\
S05  &  9.8  &  3.5  & 11.1  &  1.9  & 8.6  &  1.5 \\
S06  &  13.0  &  2.3  & 14.9  &  1.1  & 13.1  &  1.3 \\
S07  &  12.5  &  2.4  & 13.7  &  1.1  & 11.1  &  1.5 \\
S08  &  13.4  &  2.0  & 15.0  &  1.0  & 13.1  &  0.9 \\
S09  &  13.3  &  2.0  & 14.5  &  1.2  & 12.5  &  1.1 \\
S10  &  13.1  &  2.1  & 13.9  &  1.0  & 11.1  &  1.1 \\
S11  &  13.5  &  1.9  & 15.0  &  1.0  & 12.5  &  1.3 \\
S12  &  8.0  &  3.7  & 10.4  &  2.2  & 8.1  &  1.9 \\
S13  &  9.1  &  3.2  & 10.8  &  1.7  & 7.9  &  1.3 \\
S14  &  13.0  &  2.3  & 14.7  &  1.1  & 12.8  &  1.1 \\
S15  &  13.7  &  1.8  & 14.6  &  1.0  & 12.3  &  0.9 \\ \hline

    \end{tabular}
    \label{tab:gk_ages}
\end{table}

\end{appendix}

\end{document}